# *Acoustic-pressure-assisted engineering of aluminium foams*

*Xavier Mettan[†], Edoardo Martino[†], Lidia Rossi, Jaćim Jaćimović and László Forró\**

*Laboratory of Physics of Complex Matter, École Polytechnique Fédérale de Lausanne, 1015 Lausanne, Switzerland*

*+ABB Corporate Research Center, 5400 Baden-Daetwill, Switzerland*

*Juraj Krsnik, Osor S. Barišić*

*Institute of Physics, 1000 Zagreb, Croatia*

*Norbert Babcsán*

*Innobay Hungary Ltd., Miskolc Hungary*

*Sándor Beke*

*Aluinvent Zrt. Felsőzsolca Hungary*

*Rajmund Mokso*

*Paul Scherrer Institut, Villigen, Switzerland and MAX IV Laboratory, Lund University, 221 00 Lund, Sweden*

*George Kaptay\**

*Bay Zoltan Ltd for Applied Research, 2 Iglói, Miskolc, 3519, Hungary and University of Miskolc, Egyetemváros 3515, Hungary*

# Abstract

Foaming metals modulates their physical properties, enabling attractive applications where lightweight, low thermal conductivity or acoustic isolation are desirable. Adjusting the size of the bubbles in the foams is particularly relevant for targeted applications. Here we provide a method with a detailed theoretical understanding how to tune the size of the bubbles in aluminium melts *in-situ* via acoustic pressure. Our description is in full agreement with the high-rate three-dimensional X-Ray radioscopy of the bubble formation. We complement our study with the intriguing results on the effect of foaming on electrical resistivity, Seebeck coefficient and thermal conductivity from cryogenic to room temperature. Compared to bulk materials the investigated foam shows an enhancement in the thermoelectric figure of merit. These results herald promising application of foaming in thermoelectrics materials and devices for thermal energy conversion.

# Introduction

A metal foam is an arrangement formed of gas cells trapped in a solid-metal matrix, and it presents two tangible-technological advantages over its bulk counterpart. First, by tuning the size, volume fraction and shape of the bubbles, the composite mechanical, electronic and thermal properties can be tailored according to the desired function of the resulting material [1,2]. In many cases, this leads to a blend of the original material's properties with new features, augmented by the lightweight. The latter doubly profits by lowering the imprint of the foam parts in a mechanical assembly and reducing the amount of raw material, hence diminishing its cost, which is also a strategic requirement [3].

Applications of metal foams are versatile, because they combine attractive features of metals, such as high electrical and thermal conductivity and good mechanical properties with light-weight and large surface-to-volume ratios [4–6]. Amongst many existing applications, different metal-matrix-foam materials serve as heat exchangers, bodyworks, sound absorbers, electromagnetic shielding elements, catalysts in various sectors such as automotive, buildings or aerospace [7,8]. Aluminium alloys stand out as the most popular matrices for metal foams. Abundance of aluminium on Earth, its low density, good mechanical properties, and resistance to corrosion make these alloys good candidates for foaming. In addition, aluminium ranks as a low-melting-point metal, rendering its processing easier, demanding less power compared to, e.g. titanium or steel.

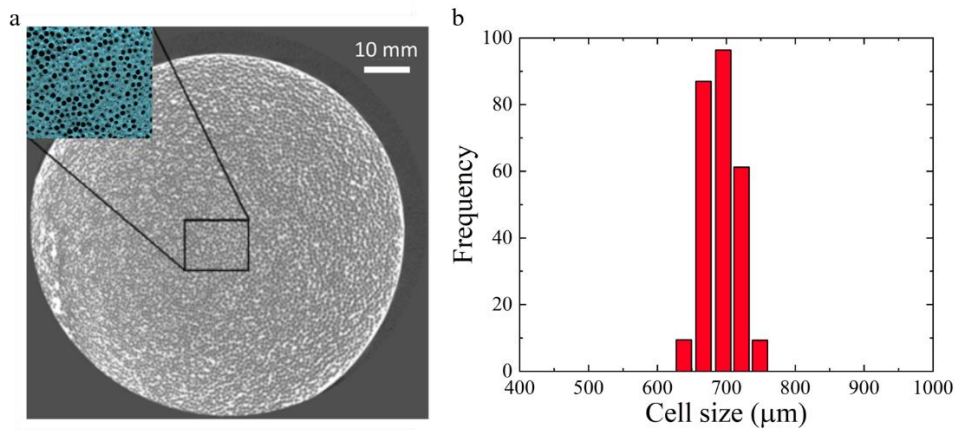

Figure 1. a. Microscopy image of a Duralcan aluminium foam prepared by acoustic pressure assistance. b. The histogram based on the image analysis reveals a reasonably uniform bubble sizes.

Due to this large interest in metal foams, several methods for foaming have been developed. That is, numerous processes have been demonstrated to achieve foaming of aluminium alloys, such as direct gas injection methods, preparation from metallic powders containing a blowing agent (a convenient one is $TiH_2$ [9]) or infiltration of the matrix into spaceholders. The latter permits a rather precise control of the cells geometry and spacing, however only open-cell configurations are achievable and the process involves many complex steps. Production from metallic powders is easier to implement, but prevents a precise and low-cost tailoring of the cells dimensions [10].

A direct injection of a gas into liquid aluminium is a well-known and straightforward route to create bubbles, where control on the shape of the foam can be exerted via injection-nozzle diameter and nozzle-wetting characteristics. Precisely controlling the size and distribution of the cells is crucial when optimising the properties of a foam for a targeted application. However, while tuning of the foam morphology is possible by adapting the hardware or changing the composition of an alloy, none of the established foaming techniques can yet allow for direct, in-situ control of the foam morphology. Providing a way to spatially engineer the distribution of cell's sizes would be beneficial for targeted applications in particular to enhance the structural strength of mechanical parts.

Here, we give a theoretical description of an acoustic-enhanced control of the size of the bubbles in an aluminium melt. A representative aluminium sample engineered with acoustic pressure is displayed in figure 1, showing a rather uniform bubble size distribution. To further verifying our model to control the foam parameters, we have experimentally investigated the process by means of synchrotron based X-Ray imaging yielding the bubble size distribution as a function of the experimental parameters. The beauty of our method is that it is not restricted to aluminium melts. Instead, it could be applied to a variety of other systems. In fact, the only requirement is to have a foamable phase with cell walls stable against external perturbations. Knowing the density of the foamable liquid and the diameter of the injector permits an estimate of the span of bubble diameter reachable by this method.

In addition, for one type of aluminium foam, we have performed the characterization of electrical and thermal transport coefficients: electrical resistivity ($\rho$), Seebeck (S) coefficient and thermal conductivity ($\kappa$) in a broad temperatures range. Our results suggest that material foaming could be a good strategy for application in thermoelectric energy conversion. Thanks to the fine tuning of all transport coefficient and enhancement of the Seebeck coefficient we have observed an improvement in the efficiency for thermal to electrical energy conversion, quantified by the thermoelectric figure of merit *ZT*. The promising target alloys could be topological insulators and Heusler alloys [11].

# Materials and methods

**Materials**

In our study, we focus on a commercially available aluminium metal-matrix composite (MMC) called Duralcan F3S.20S, that contains 20% in volume of SiC particles of 20 µm diameter (Chemical composition of F3S.20S in %wt.: Si:9.2, Fe:0.12, Mn:0.02, Mg:0.54, Ti:0.10, Al:balance, SiC:20.8). These solid particles are known to stabilise the foams in the liquid state thanks to the action of interfacial capillary forces [12].

Foams made by these metal-matrix composites also have improved stability and can be manipulated and solidified to form any shape [13, 14]. They are popular in industry owing to their low density and good mechanical properties.

**Methods**

Samples of metal foams were fabricated according to the method presented in Ref [15], as depicted in Figure 2a. Liquid Duralcan was held in a heating crucible at a temperature of 700 °C. Gas (argon 4.6) was injected at controlled and regulated (PID) flow at a pressure of 2.5 bar through the melt via a nozzle coupled to an ultrasonic-wave generator, called sonotrode, placed at the bottom of the crucible, and with its longitudinal axis pointing upwards. The orifice of the sonotrode has a diameter $D_i = 0.2$ mm and the external diameter of the circular sonotrode is $D_e = 4$ mm. A 30 kHz ultrasonic generator (Woodpecker UDS-N1) delivering up to 160 W/cm2 of acoustic power and a piezo-electric transducer (titanium waveguide) were employed to pulse acoustic waves through a steel injector, so that the injector vibrates along its longitudinal axis (vertically). The characteristic bubble size $D(P_{el})$ was measured as the function of the control parameter $P_{el}$, being the electrical power (0 - 20W) fed to the sonotrode [9]. In addition, the differential pressure prevailing at the orifice of the nozzle, called dynamic pressure, was monitored as a function of time. A differential-pressure sensor (Motorola MPX5050DP) connected to the steel injector provides a voltage signal to custom acquisition electronics, with 5 Pa accuracy with 500 khz sampling rate. In a steady-state regime, when bubbles form and detach at a constant rate, the dynamic pressure oscillates with a period corresponding to the time of formation of a bubble $t_b$. Knowing the constant gas-flow rate $Q$, the volume of the bubble is determined as $V_b = Q \cdot t_b$. The broad range of bubble size as a function of acoustic pressure, translated to sonotrode power is used to support the theoretical description.

High-brilliance and high-coherence X-Ray beam (TOMCAT beamline at the Swiss Light Source, Paul Scherrer Institute) allows micro- and sub-micro-metre, quantitative, three-dimensional imaging at high frame-rate and extend the traditional absorption imaging technique to edge-enhanced and phase-sensitive measurements. In the current study the probe was a broad-band X-ray beam with energy centred around 20 keV (wavelength : 0.06 nm). The foaming furnace was placed in the X-ray beam in front of a 500 μm-thick YAG:Ce scintillator coupled to a high-speed camera (PCO Dimax) optimized for sub-millisecond recording. The formation of bubbles in the melt was monitored by pressure measurement simultaneously with the X-ray radioscopy at 1400 fps, 6600 fps and 10'000 fps with pixel size of 11 μm. In order to determine the diameter of bubbles, still images were analysed with ImageJ v.1.43u with 2% accuracy; the diameter of the nozzle was employed as a scale reference (rectangle highlighted in Figures 2b and 2c). Such measurements of diameters were repeated for at least 10 times, from two perpendicular angles of view with respect to the nozzle, yielding a statistical average for diameters at various acoustic pressures. The size of bubbles was adapted by the acoustic pressure, generated by the sonotrode in the liquid metal (Figure 2b and c).

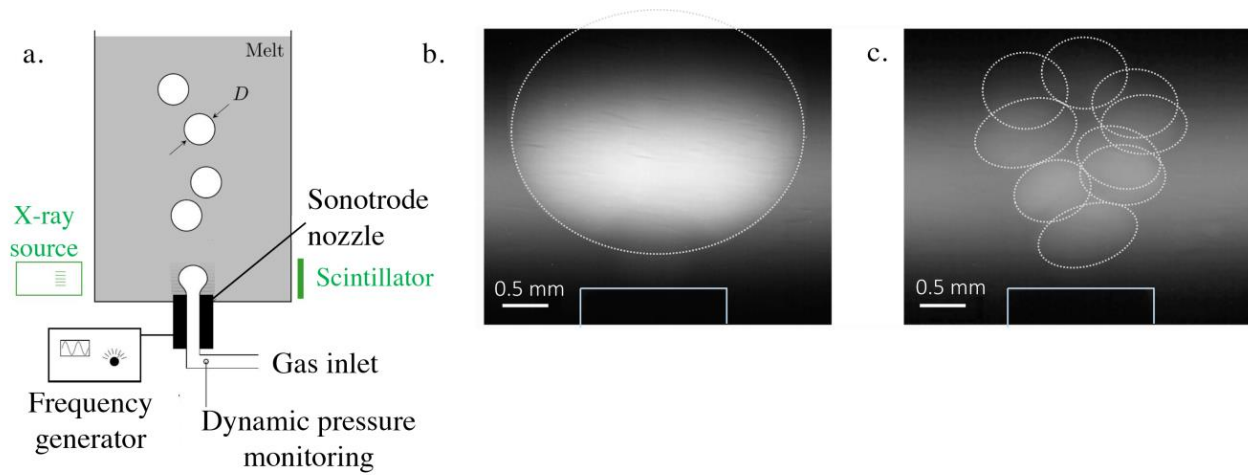

**Figure 2. a**. Sketch of the experimental setup for investigating foaming process by *in-situ* X-Ray radioscopy imaging of the bubbles. A sonotrode nozzle injects the gas inside the liquid (e.g. argon into aluminium melt). The control on the acoustic-radiation pressure is ensured by a frequency generator, controlling the transducer of the sonotrode. For a given gas-flow rate, the bubbles have a characteristic diameter *D*; **b.** and **c.** are the characteristic X-ray snapshots of the bubbles in molten Duralcan with two values of electrical power fed to the sonotrode. The contour of the bubbles and the nozzle are highlighted for better visibility.

# Acoustic-pressure-induced detachment of gas bubbles in a liquid

Here we assess the influence of acoustic pressure on the size of bubbles in a liquid, and we introduce a general model for their acoustic-modulation, which can be applied to different liquids. First, a common model for determining the size of bubbles in a liquid under static-pressure conditions is recalled, after which step we introduce an effective contribution for dynamic modulation. We consider a situation with a cylindrical nozzle, pointing upwards and submerged in a liquid (Figure 3). The nozzle can inject gas in the liquid at a controlled-flow rate and pressure. Typically, at low flow rates, a bubble starts to form at the orifice and grows to a special shape close to that of a sphere before it detaches from the orifice.

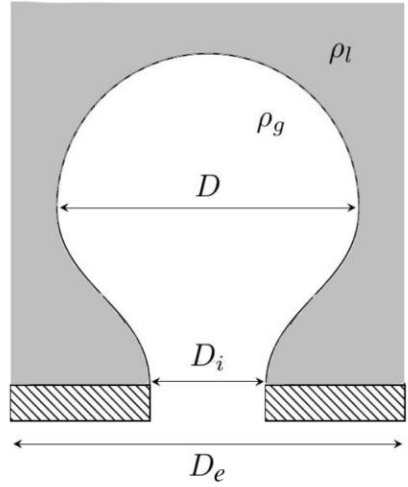

Figure 3. Sketch of the formation of a gas bubble (density $\rho_g$) inside a the liquid metal ($\rho_l$) with the parameters: $D_e$ is the external diameter of the sonotorode, $D_i$ the inner diameter of the nozzle, and the bubble neck, while $D$ is the maximum diameter perpendicular to the gas inlet.

In figure 3, an idealised situation, just before detachment of the bubble is depicted. The bubble is considered symmetrical around the vertical axis parallel to the nozzle [16]. At this instant, a simple force balance allows us to predict the characteristic diameter $D$ of the bubble after detachment, supposing it has a spherical shape. First the buoyancy force $F_b$, pointing upwards, tends to detach the gaseous bubble from the nozzle:

$$F_b = g(\rho_l - \rho_g)V_b, \qquad (1)$$

where $V_b$ is the volume of the bubble, $g = 9.81$ m/s² the gravitational acceleration and $\rho_l$ ($\rho_g$) is the density of the liquid (gas). The volume of the bubble is modelled through its diameter just after it detaches from the orifice. As $\rho_g \ll \rho_l$, the buoyancy force in Eq. (1) is simplified as $F_b = \frac{\pi}{6}\rho_l g D^3$. Opposed to $F_b$, the interfacial anti-stretching force acting at the neck of the bubble is modelled as [17]:

$$F_t = -L_i \gamma, \qquad (2)$$

where $\gamma$ is the surface tension of the liquid, $L_i = \pi D_i$ is the inner perimeter of the orifice measured along the horizontal plane, perpendicular to the direction of the stretching buoyancy force [18].

It is appropriate to consider the inner perimeter $D_i$ when the liquid wets the material of the orifice. Writing $F_b + F_t \geq 0$ (at the instant of detachment), the characteristic diameter of a bubble can be obtained by substituting Eq. (2) into the latter condition:

$$D^0 \geq \left(\frac{6\gamma D_i}{g\rho_l}\right)^{1/3} \qquad (3)$$

which can be verified experimentally. The superscript "0" denotes the absence of any additional external forces. If one inserts $\rho_l = 2300$ kgm⁻³, $\gamma = 0.86$ Jm⁻² of oxidized liquid aluminium into Eq. (3) with the applied inner nozzle diameter of 0.20 mm, the value of $D^0 = 3.6$ mm is obtained. This is the boundary condition (maximum value) of the detaching bubble size from this nozzle in case of no pressure applied from the sonotrode. This value is in agreement with data at zero pressure, shown in figure 4b.

One can notice that in Eq. (3) for a given liquid, only $D_i$ enables the control of the size of the bubbles. However, in applications this quantity cannot be tuned *in-situ*, especially in challenging conditions of high-temperature melts. In order to overcome these technological limitations, we introduce an additional force acting on the bubble, promoted by acoustic pressure. Conveniently, the submerged orifice can be a nozzle combined with a sonotrode device (see figure 2), which vibrating power and frequency can be electronically actuated. As a first approach, we consider an additional effective force $F_a$ opposing the interfacial anti-stretching force, so that the new condition for the bubble to detach from the orifice is: $F_b + F_t + F_a \geq 0$. The force $F_a$ applies on the projection of the surface of the bubble not covered by the nozzle $A_p = \pi(D^2 - D_i^2)/4$ and can be written as:

$$F_a = p \cdot A_p = p(P_{el}) \cdot \pi[D^2(P_{el}) - D_i^2]/4, \tag{4}$$

where $P_{el}$ is the electrical power of the frequency generator supplied to the sonotrode via the piezo-electric transducer, $p = p(P_{el})$ is the acoustic pressure and $D = D(P_{el})$ is the control parameter dependent bubble diameter. Finally, the balance equation for the bubble diameter $D$ in a nozzle-wetting liquid is:

$$g\rho_l D^3 + \frac{3}{2}p(D^2 - D_i^2) - 6D_i\gamma = 0. \tag{5}$$

Eq. (3) is obtained back from Eq.(5), fulfilling the boundary condition: at $p(0) = 0$, $D(0) = D^0$. A formulation of $p(P_{el})$ can be made recalling simple concepts. The actual power density transmitted to the liquid is $I_s = \eta P_{el}$, where $0.1 < \eta < 1$ is a transmission efficiency and $A_s = \frac{\pi}{4}(D_e^2 - D_i^2)$ is the area of the sonotrode in contact with the liquid and perpendicular to its oscillation direction. The actual pressure reflected [7] on the liquid-gas interface is:

$$p_{ar} = 2\frac{I_s}{v_s} = 2\eta\frac{P_{el}}{v_s}. \tag{6}$$

Here, $v_s$ is the speed of sound in the liquid medium and the factor 2 stems from the complete reflection of the sound wave at the liquid-gas interface [19 - 21].

Finally, the acoustic pressure $p$ can be written as:

$$p(P_{el}) = k_{\text{eff}} \cdot p_{ar} = k_{\text{eff}} \cdot \frac{1}{v_s} \cdot \frac{8\eta}{\pi(D^2 - D_i^2)} \cdot P_{el} = b \cdot P_{el}, \tag{7}$$

where the only dependence on the liquid type is in $v_s$. The parameter $k_{\text{eff}}$ accounts for damping inside the liquid [22]. For example, for a sphere with large Reynolds number, $k_{\text{eff}} \simeq 0.4$ [23].

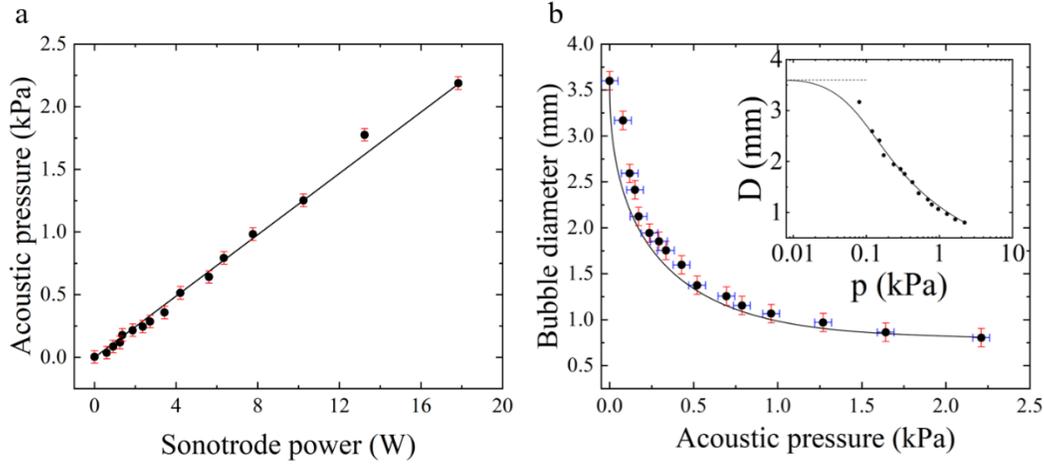

Figure 4 a. Showing the linearity of $p(P_{el})$ with $P_{el}$, in the spirit of the model of acoustic pressure. The solid line corresponds to Eq. 7, with $b = 2.5$ Pa/W; b. Collected data of bubble diameter as a function of the acoustic pressure (see figure 2). The solid line is a fit according to Eq. (5). Since some of the bubbles have oval shape, the size was measured in two perpendicular directions and the average value was attributed to D. This sets the error bar of the data. The inset shows the same data in a semi-logarithmic representation, to highlight the presence of an inflexion point in the curves, where the buoyancy force dominated bubble formation is overtaken by the acoustic pressure.

As seen in figure 4a, the acoustic pressure can be modelled by the form proposed in Eq. (7). In this manner, the fitting parameter $b$ provides an estimation for the diameter of the bubble as a function of $p = bP_{el}$. The bubble diameter $D$ as a function of the acoustic pressure $p$ is shown in figure 4b for Duralcan. The fitting line for Duralcan fairly agrees with the experiment.

Here we have shown experimentally and theoretically that applying ultrasonic power to the melt permitted us to control the diameter of bubbles continuously from 3.8 to 0.7 mm. It has significant implications for fabrication of aluminium foams. First, one can prepare products with customised pore sizes on the same production line, which has obvious cost advantages. Second, materials with custom bubble-diameter patterns can be designed. For example, one could imagine large foam panels with gradient pore diameter, with pores smaller and smaller as they become closer to the surface, a geometry that would favour light-weight and good strength. This could even be more important for sound absorption, where an accurate-controlled distribution of the pores would lead to specific acoustic properties. Thermal-flow control could also be enhanced through alternation of large/small pores.

# Measurement of transport coefficients

A novel possibility for application of metal foaming could be in thermoelectricity, which is a direct conversion of the waste heat into electricity. As a proof of principle demonstration that foaming could increase the figure of merit *ZT* of thermoelectric devices, we have measured transport properties of the solidified Duralcan foam with the porosity 76%, and made a comparison to the bulk parent material (before foaming), to the compressed foam, and to the 5N pure aluminium. In particular, we have found that the foaming increases ZT by an order of magnitude around room temperatures, ascribing this effect to the enhancement of thermopower due to the reduced dimensionality in a disordered environment and electron filtering effects.

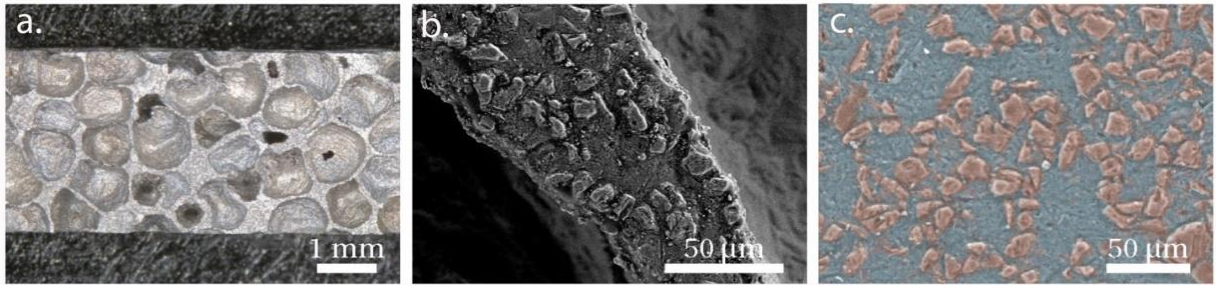

Figure 5. Optical and SEM images, and compositional analysis (EDX) of the Duralcan composite with 20% in mass of SiC used for bubble-formation and transport coefficient measurements. a) Optical micrographs of the specimen shaped for transport measurements; b) SEM images of the matrix; and c) EDX color maps of the matrix. Light blue corresponds to aluminium, pale orange to silicon, corresponding to SiC grains. The detailed characterization of the foam is: Al: 86.3 %wt, Si: 13.2 %wt, Mg: 0.3 %wt; O < 1.0 %wt; SiC 20 %wt, and size distribution 10-20 μm; bubble size 0.7±0.1 mm and porosity 76%.

The measured samples of foam were cut from a massive block to a typical dimension of 1 x 3 x 10 mm$^3$. The structure and the compositional analysis are depicted in figure 5. The optical and SEM images reveal the textured matrices. It can be seen that the sample has closed-cell pores of average diameter between 700 μm, however their distribution varies somewhat. The SEM and EDX images clearly show the presence of SiC particles as surfactant, stabilizing the bubbles. Their size is in the 10-20 μm range. One may notice a small quantity of Mg as well, which is a constituent of the Al alloy matrix.

The transport coefficients investigated here are the electrical resistivity ρ, the thermopower S, and the thermal conductivity κ. These three coefficients have never been studied for the same kind of aluminium foam, and especially not in a broad temperature range involving liquid helium temperatures, which is important for a microscopic understanding of transport properties. The thermoelectric figure of merit is given by,

$$ZT = \frac{\sigma S^2 T}{\kappa} \qquad (8)$$

where T is the temperature and $\sigma = 1/\rho$ is the electrical conductivity. The *ZT* value, above which the material is suitable for thermoelectric applications, is 1. Figures 6a, b and c show the temperature dependence of ρ, κ, and S, respectively, for high purity aluminium (5N), the bulk Duralcan, the Duralcan foam, and the sample of compressed Duralcan foam where the density is close to the solid composite. To obtain the latter, a small block of foam has been uniaxial compressed with pressure up to 0.8 GPa at 200°C to completely collapse all the bubbles.

The resistivity ρ(T) for the pure aluminium in figure 6a closely reproduces results found in the literature [24]. For the Duralcan, by using the simple Drude model, the residual resistivity ρ (T≈0) provides an estimate of the electron mean-free path of approximately 20 - 30 nm. The latter roughly matches the average crystallites size (30 - 40 nm). With foaming, the absolute value of ρ (T≈0) is shifted to the 60 - 70 μΩcm range. The high value of ρ(T) for the foam indicates that in addition to the scattering on lattice vibrations, scattering at bubbles' surface, at interconnections, and on the SiC particles strongly alter the electronic mean free path. Interestingly, when the residual resistivity is subtracted from the data, the temperature-dependence of the resistivity measurements for all samples may be satisfactory described by the Bloch-Grüneisen formula, with the Debye temperature of roughly 400 K.

The Wiedemann-Franz law states that $\kappa_{el}$ is proportional to the electrical conductivity, $\kappa_{el} = L_0 T \sigma$, where $L_0 = 2.44 \cdot 10^{-8}$ WΩK$^{-2}$ is the Lorenz number. Using this law, the thermal transport for the pure aluminium may be almost entirely ascribed to the electron subsystem, $\kappa \approx \kappa_{el}$. While the high values of the thermal conductivity $\kappa$ for the pure aluminium in figure 6c are characteristic for crystalline metals, $\kappa$ for the Duralcan samples collapses to low values, with a full suppression of the maxima at low temperatures that characterizes $\kappa$ for the pure aluminium. This kind of behavior of $\kappa$, similarly as in the case of resistivity data, suggests that electrons are strongly scattered. By extracting through the Wiedemann-Franz law the electronic contribution from the total $\kappa = \kappa_{el} + \kappa_{ph}$, one could extract the phonon contribution $\kappa_{ph}$. Such kind of an analysis indicates that the relative phonon contribution to $\kappa$ is significant only for the foam and comparable to $\kappa_{el}$ at elevated temperatures.

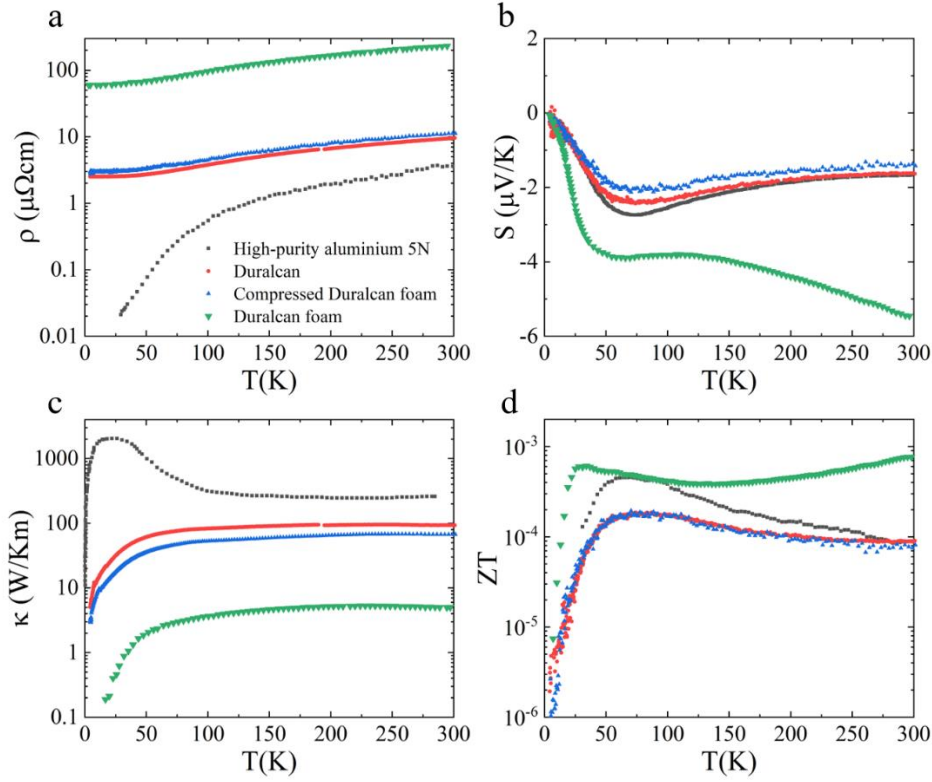

Figure 6. Comparison of temperature-dependent transport coefficients and figure of merit (*ZT*) for: high-purity 5N aluminium, bulk Duralcan, compressed Duralcan foam and, Duralcan foam. a) Electrical resistivity $\rho$, b) thermoelectric power *S* (the curve for pure aluminium is reproduced from [26]), c) thermal conductivity κ (the curve for pure aluminium is reproduced from [25]), and d) the figure of merit *ZT*.

The results for $\sigma$ and $\kappa$ clearly show a dramatic effect that micro-structuring of the material has on transport properties [27-31]. However, with exceptions of low temperatures, the ratio of these two coefficients in Eq. (8) remains similar for all samples. Therefore, we turn to the thermoelectric power *S* for the bulk Duralcan samples and the foam, shown in figure 6c, in order to rationalize the large enhancement of *ZT* observed in figure 6d for the foam. The two depict similar temperature dependence, with the local maximum of |S| due to the significant influence of the electron-phonon coupling, although that of the latter has doubled at the room temperature. The origin of this difference may be traced back by using the Mott formula:

$$S \sim d\ln\sigma/dE = (d\ln\mu(E)/dE + d\ln N(E)/dE)_{E=E_F} \qquad (8)$$

where the first term represents the energy dependence of the electron mobility, while the second is given by the energy dependence of the density of states (DOS) [32]. We argue that foaming and the SiC inclusions may greatly enhance the second term. Reduced dimensionality/quantum confinement effects [33-36] in the foam and many impurity states introduce large variations in the local density of states, thus enhancing the derivative of local DOS with respect to the energy in the vicinity of the Fermi level. Moreover, the electron diffusion part of the thermopower may be additionally increased due to the electron filtering effects [37-39], where bubbles act as tall barriers. This is an extremely interesting development because it justifies our conjecture that foaming could be beneficial for increasing *ZT*, as shown in figure 6d, by an order of magnitude increase for the foam over that of the bulk material. It is plausible that further engineering of the foam could result by even stronger enhancements of *ZT*.

# Conclusions

We have modelled the bubble formation by applying acoustic pressure in molten metals, which fully describe the experimental observations of *in-situ* synchrotron-based X-ray radioscopy of bubbles in Duralcan aluminium composites. This method allows fine tuning of the size of the bubbles that could be stabilized by additives like SiC particles, leading to castable, highly functional metallic foams.

Performing detailed electronic and heat transport studies of the aluminium foam, we have given a proof of principle that foaming a metal could significantly enhance the thermoelectric figure of merit. The class of materials where such a strategy could be highly beneficial are topological insulators and Heusler alloys.

# Acknowledgments

GK thanks projects TÁMOP-4.2.2.A-11/1 / KONV-2012-0036 and GINOP-2.3.2-15-2016-00027 obtained from the Ministry of Innovation and Technology of Hungary. The work in Lausanne was supported by the Swiss national Science Foundation. The Technical support of Peter Makk is greatfully acknowledged. JK acknowledges the support by the Croatian Science Foundation Project IP-2016-06-7258. OSB. acknowledges the QuantiXLie Center of Excellence, a project co-financed by the Croatian Government and European Union (Grant No. KK.01.1.1.01.0004). We acknowledge the Paul Scherrer Institute for granting experimental time at the TOMCAT beamline (proposal 20100211).

† these authors contributed equally to this work

*To whom correspondence should be addressed: laszlo.forro@epfl.ch; kaptay@hotmail.com